\documentclass[preprint]{revtex4-1}

\usepackage{graphicx}
\usepackage{graphics}
\usepackage{color}
\usepackage{ulem}
\usepackage{subfig}
\usepackage{natbib}
\usepackage{amsmath}
\usepackage{amssymb}
\usepackage{placeins}
\usepackage{float}
\floatstyle{plain}
\restylefloat{figure}
\usepackage{nicefrac}
\usepackage{xfrac}

\usepackage{ulem}

\newcommand{\slfrac}[2]{\left.#1\middle/#2\right.}

\begin{document}

%\preprint{APS/123-QED}

\title{Velocity Profile in a Two-Layer Kolmogorov-Like Flow}

\author{Balachandra Suri, Jeffrey Tithof, Radford Mitchell, Jr., Roman O. Grigoriev, Michael F. Schatz}
\affiliation{Center for Nonlinear Science and School of Physics, Georgia Institute of Technology, Atlanta, Georgia 30332-0430, USA}

\date{\today}

%\pacs{}

\begin{abstract}
In this article we discuss flows in shallow, stratified horizontal layers of two immiscible fluids. The top layer is an electrolyte which is electromagnetically driven and the bottom layer is a dielectric fluid. Using a quasi-two-dimensional approximation, we derive the depth-averaged two-dimensional (2D) vorticity equation which includes a prefactor to the advection term, previously unaccounted for. In addition, we study how the horizontal components of velocity vary in the vertical direction. For a Kolmogorov-like flow, we evaluate analytical expressions for the coefficients in the generalized 2D vorticity equation, uncovering their dependence on experimental parameters. To test the accuracy of these estimates, we experimentally measure the horizontal velocity fields at the free-surface and at the electrolyte-dielectric interface using particle image velocimetry (PIV). We show that there is excellent agreement between the analytical predictions and the experimental measurements. Our analysis shows that by increasing the viscosity of the electrolyte relative to that of the dielectric, one may significantly improve the uniformity of the flow along the vertical direction.
\end{abstract}

\maketitle

\section{Introduction}

The study of two-dimensional (2D) flows has received significant attention in recent decades with the aim of understanding turbulence \citep{tabeling_2002}. Compared to their three-dimensional (3D) counterparts, 2D flows are more amenable to theoretical analysis.  However, flows in the real world are never strictly 2D; one may only approach two-dimensionality as the components of velocity parallel to a plane (horizontal) become much stronger than the one perpendicular to it (vertical). Such flows have often been referred to as quasi-two-dimensional (Q2D) flows \citep{dolzhanskii_1992}. Q2D flows have been experimentally realized in a variety of systems which include flows in shallow  electrolytic layers \citep{bondarenko_1979}, superfluid helium \citep{campbell_1979}, liquid metals \citep{sommeria_1982},  soap films \citep{couder_1984}, and electron plasmas \citep{mitchell_1993}. The mechanism that leads to suppression of the vertical velocity component is different in each of these systems, demanding a specialized approach. In this article, we discuss flows in shallow electrolytic layers, a system which has been subject to extensive study due to ease of experimental realization.

Fluid flows in shallow electrolytic layers have been realized experimentally in homogeneous \citep{bondarenko_1979} as well as stratified layers of fluids \citep{ cardoso_1994,rivera_2005}. It was first observed by \citet{bondarenko_1979}, for an experimental realization of the Kolmogorov flow in a homogeneous electrolytic layer, that the interaction of the flow with the solid boundary at the bottom resulted in dissipation that was not accounted for in the 2D Navier-Stokes equation (NSE). It was suggested that the addition of a linear term ($-\alpha\bf{u}$) to the 2D NSE modeled the dissipation satisfactorily \citep{bondarenko_1979}; this term is commonly referred to as the ``Rayleigh friction" term. The stability of the laminar flow predicted by the 2D NSE with friction (NSE-WF) was in good agreement with the one experimentally observed.  A more thorough discussion of related theoretical and experimental results is provided in the articles by \citet{obukhov_1983} and \citet{thess_1992}.

The question as to how accurately the 2D NSE-WF describes flows realized in experiments attracted significant attention in the mid-1990s. Measuring the horizontal velocity field of a decaying dipolar vortex in an electrolyte layer, \citet{paret_1997b} inferred that the horizontal velocity, following a brief transient, relaxed to a plane Poiseuille profile in the vertical direction. The measured rate of decay was also in good agreement with that of plane Poiseuille flow \citep{rivlin_1984}. \citet{juttner_1997} performed a DNS of the 2D NSE-WF, using experimental data to initialize the simulation, and showed that it did capture the evolution of the decaying flow, at least qualitatively. Later, \citet{satijn_2001} analyzed the decay of a monopolar vortex using a full 3D DNS and reported a regime diagram showing that a weakly driven flow in shallow electrolytic layers (both homogeneous and stratified) remained Q2D during its decay. Following this study, Akkermans {\textit{et al.}} performed experimental measurements of 3D velocity fields using Stereo-PIV (SPIV) in a single layer setup \citep{akkermans_2008a,akkermans_2008b} and a two-layer setup \citep{akkermans_2010} at high Reynolds numbers ($Re \approx 2000$). Their results indicated that, at these Reynolds numbers, the vertical velocity component was comparable to the horizontal components and hence the flow could no longer be considered Q2D. To understand the transition of a Q2D flow to a 3D one, \citet{kelley_2011a} performed experiments over a wide range of Reynolds numbers ($30< Re <250$) and showed that there is a critical Reynolds number ($Re_c \approx 200$), both for homogeneous and stratified layers, below which the flow can be considered Q2D. These studies, aimed at understanding the three-dimensionality of flows in shallow electrolytic layers, suggest that there are three mechanisms that lead to the onset of three-dimensionality. Eckman pumping which results from the variation of vorticity with depth \citep{akkermans_2008b, kelley_2011a} and interfacial deformations which drive gravity waves \citep{akkermans_2010} are in play at all Reynolds numbers. Shear instability, on the other hand, sets in above a critical Reynolds number \citep{kelley_2011a}.

Most experiments studying flows in shallow electrolytic layers were aimed at understanding 2D turbulence from a statistical perspective, requiring high Reynolds numbers to be realized \citep{cardoso_1994, marteau_1995, paret_1997b}. However, in recent years, there has been moderate success, both on theoretical \citep{nagata_1997,kerswell_2005,eckhardt_2007,gibson_2009} and experimental fronts \citep{hof_2004,lozar_2012}, in understanding transitional and weak turbulence (in both 2D and 3D) as dynamics guided by exact but unstable solutions (often referred to as Exact Coherent States (ECS)) of the NSE. In 2D, for instance, \citet{chandler_2012} have recently identified around 50 different ECS at low Reynolds numbers (Re $\approx$ 40) in a 2D DNS of turbulent Kolmogorov flow. This is a very significant result, since experimental flows at this Reynolds number can be considered Q2D \citep{kelley_2011a, satijn_2001}. However, aside from the study of \citet{figuearoa_2009}, to the best of our knowledge, there have been no attempts at making a quantitative comparison between experiments and numerical simulations of forced flows in shallow electrolytic layers. Hence, building a framework for a direct comparison between a Q2D flow and a 2D model used to describe such a flow is imperative. In particular, such a framework is necessary to describe how the experimental parameters -- fluid layer depths, viscosity, density, and the forcing -- affect the flow.

In this article, we derive a generalized 2D vorticity equation, describing the evolution of weakly driven flows in shallow electrolytic layers. For two special cases, the Kolmogorov flow and unidirectional flow, we obtain the profile of the horizontal velocity field along the vertical direction for a setup with two immiscible fluid layers. Using this profile, we evaluate analytical expressions for the coefficients that appear in the generalized 2D vorticity equation. The theoretical profile is validated by measuring the velocity at the free-surface and at the interface of the two layers. Finally, we define a measure of two-dimensionality in the (forced) upper layer and show that it is possible to make the flow in that layer essentially 2D by increasing its viscosity relative to the (lubricating) lower layer.
 
\section{Generalized 2D Vorticity Equation}\label{sec:model} 

Consider a shallow layer of fluid, with thickness $h$ in a laterally extended container with a flat bottom. We assume the $xy$-plane is parallel to the bottom of the container and the $z$-axis is in the vertical direction, with $z=0$ chosen at the bottom of the fluid layer. The velocity field in such flows is inherently three-dimensional, in the sense that it generally depends on all three coordinates, ${\bf V}={\bf V}(x,y,z,t)$. This inherent three-dimensionality is due to the fact that the bottom of the fluid layer ($z = 0$) is constrained to be at rest due to the non-slip boundary condition. The velocity field in such a system is governed by the Navier-Stokes equation for an incompressible fluid ($\nabla \cdot \bf{V} = 0$)

\begin{equation}\label{eq:ns3}
\rho(\partial_t\bf{V} + {\bf{V}}\cdot{\bf{\nabla}}{\bf{V}}) = -\nabla p + \mu \nabla^2 {\bf{V}} + {\bf{f}} + \rho{\bf{g}},\\
\end{equation}
where $\rho$ is the density of the fluid under consideration, $\mu$ is the dynamic viscosity$, \rho\bf{g}$ is the gravitational force (along the $z$-axis), and $\bf{f}$ is the electromagnetic force in the plane of the fluid (the $xy$-plane). For stratified layers, the quantities $\rho$, $\mu$, and {\bf{f}} would be defined appropriately within each layer. 

For shallow layers of fluids driven by a weak, in-plane forcing, the velocity component along the vertical direction is negligible compared to the horizontal ones. For flows with the characteristic horizontal length scale $L$ which is substantially larger than the thickness $h$, the characteristic times describing equilibration of momentum in the vertical direction ($\rho h^2/\mu$) are much smaller than those associated with the horizontal directions ($\rho L^2/\mu$). Furthermore, if the direction of the forcing {\bf{f}} is independent of $z$, we can assume the {\it direction} of the velocity to be independent of the height $z$, allowing the velocity field to be factored as \citep{dovzhenko_1981, figuearoa_2009} 
\begin{equation}\label{eq:2dtype}
{\bf V}(x,y,z,t) = P(z){\bf{U}}(x,y,t) \equiv P(z) [u_x(x,y,t){\bf{\hat{x}}} + u_y(x,y,t){\bf{\hat{y}}}],
\end{equation}
where $P(z)$ describes the dependence of the horizontal velocity on $z$. The no-slip boundary condition at the bottom of the fluid layer is imposed by chosing $P(0) = 0$. Furthermore, we impose the normalization condition
\begin{equation}\label{eq:norm}
P(h)=1
\end{equation}
to make the factorization unique, so ${\bf{U}}(x,y,t)$ can be interpreted as the velocity of the free surface ($z=h$).

Substitution of (\ref{eq:2dtype}) into (\ref{eq:ns3}) gives
\begin{eqnarray}\label{eq:genprof}
\rho P\partial_t{\bf U} +  \rho P^2{\bf{U}}\cdot{\bf{\nabla}}_{\parallel}{\bf{U}} &=& - {\bf{\nabla_{\parallel}}}p + P\mu \nabla^2_{\parallel} {\bf{U}} +{\bf{U}}\mu \nabla^2_{\bot}P  + {\bf{f}},\\
{\bf{\nabla_{\bot}}}p &=& \rho{\bf{g}}, \nonumber
\end{eqnarray}
along with ${\bf{\nabla_{\parallel}}}\cdot{\bf{U}} = 0$, where the subscripts $\parallel$ and $\bot$ represent the horizontal and vertical components, respectively. In general, the profile $P(z)$ depends on the exact form of forcing ${\bf f}$ and the horizontal flow profile ${\bf U}$. However, we further assume that the profile $P(z)$ is independent of ${\bf{U}}$. This assumption, though not intuitive, proves to be valid at moderate $Re$, as we shall show for a couple of test cases below.

Integrating the first of the two equations in (\ref{eq:genprof}) over the $z$ coordinate, i.e., from the bottom of the fluid layer ($z=0$) to the free surface ($z=h$),  and taking a curl, we obtain an equation for the vertical component of the vorticity $\Omega=\partial_xu_y-\partial_yu_x$:
\begin{equation}\label{eq:2dvor}
\partial_t\Omega + \beta{\bf U}\cdot{\bf{\nabla}}_{\parallel}{\Omega}=
\nu \nabla^2_\parallel \Omega - \alpha\Omega + W,
\end{equation}
where the parameters $\beta$, $\nu$, and $\alpha$ are defined as follows

\begin{equation}\label{eq:2dterms}
\begin{aligned}
\beta = \frac{\int_0^{h}\rho P^2 dz}{\int_0^{h}\rho P dz}, \quad \nu = \frac{\int_0^{h}\mu P dz}{\int_0^{h}\rho P dz},  \quad &\alpha = \frac{\mu P'(0)}{\int_0^{h}\rho P dz}.\\
\end{aligned}
\end{equation}
The source term $W$ on the right-hand side of (\ref{eq:2dvor}) corresponds to the $z$-component of the curl of the depth averaged force density
\begin{equation}
W = \frac{\int_0^{h_d+h_c} {(\partial_x f_y - \partial_y f_x)} dz}{\int_0^{h_d+h_c}\rho P dz}.
\end{equation}

It must be noted that the prefactor for the advection term, computed for the Poiseuille profile \citep{satijn_2001} which has traditionally been used to describe Q2D flows turns out to be $\beta=\pi/4=0.78$, which is significantly different from unity, as assumed in all previous studies that used this equation. Furthermore, we recover the expression for the Rayleigh friction coefficient $\alpha=\pi^2 \nu/4 h^2$ for a Poiseuille profile \citep{juttner_1997,satijn_2001}, without assuming a decaying flow.

It is important to point out that the vorticity equation (\ref{eq:2dvor}) is a 2D equation that {\it{quantitatively}} describes 3D flows in regimes where ansatz (\ref{eq:2dtype}) is valid. In particular, $\Omega$ describes the vorticity at the top surface of the electrolyte, facilitating direct comparison between experiment and analytical or numerical solutions.

\section{Experiment}

\begin{figure}
\centerline{\subfloat[]{\includegraphics[width=2.5in]{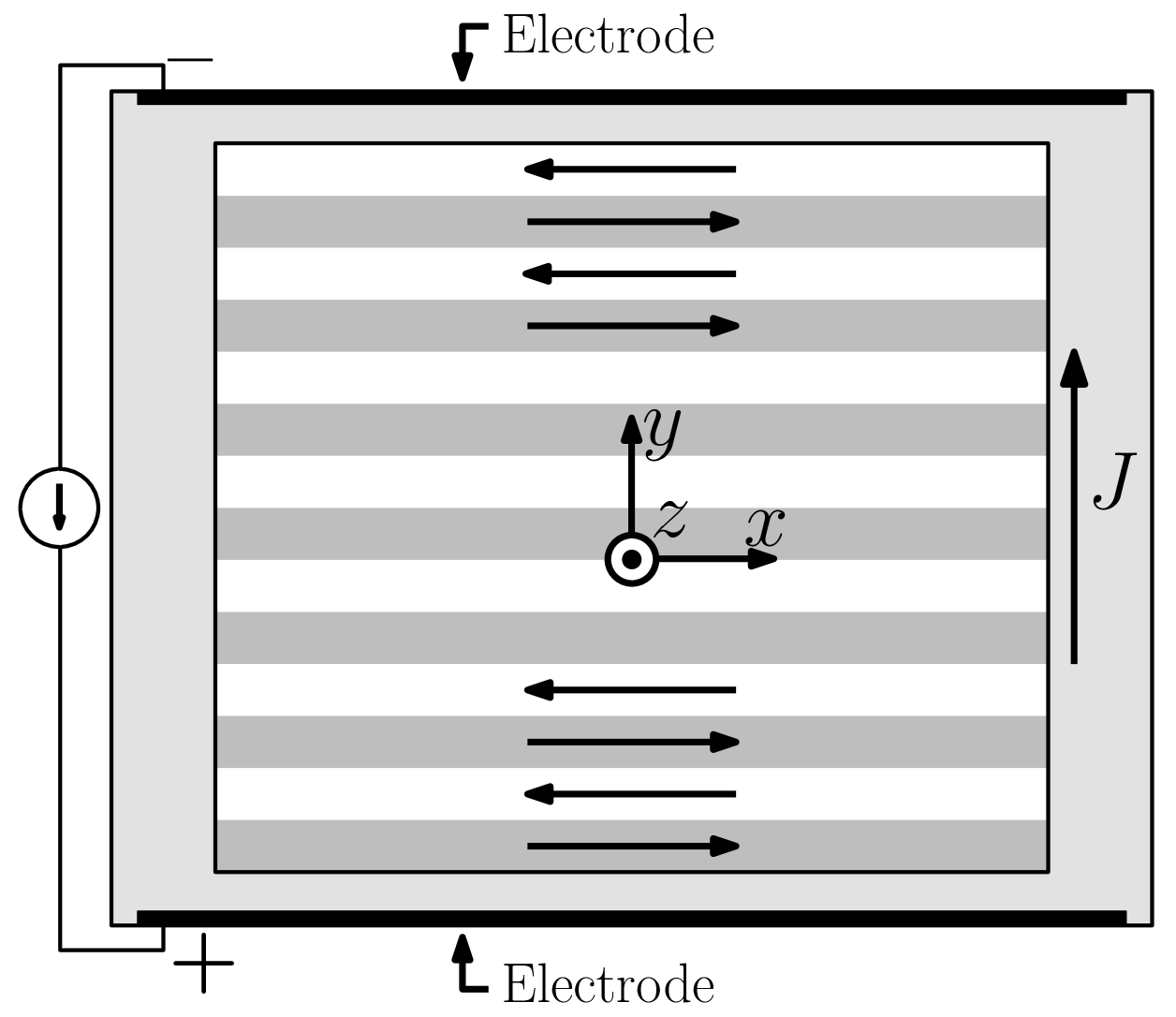}}
\subfloat[]{\includegraphics[width=2.5in]{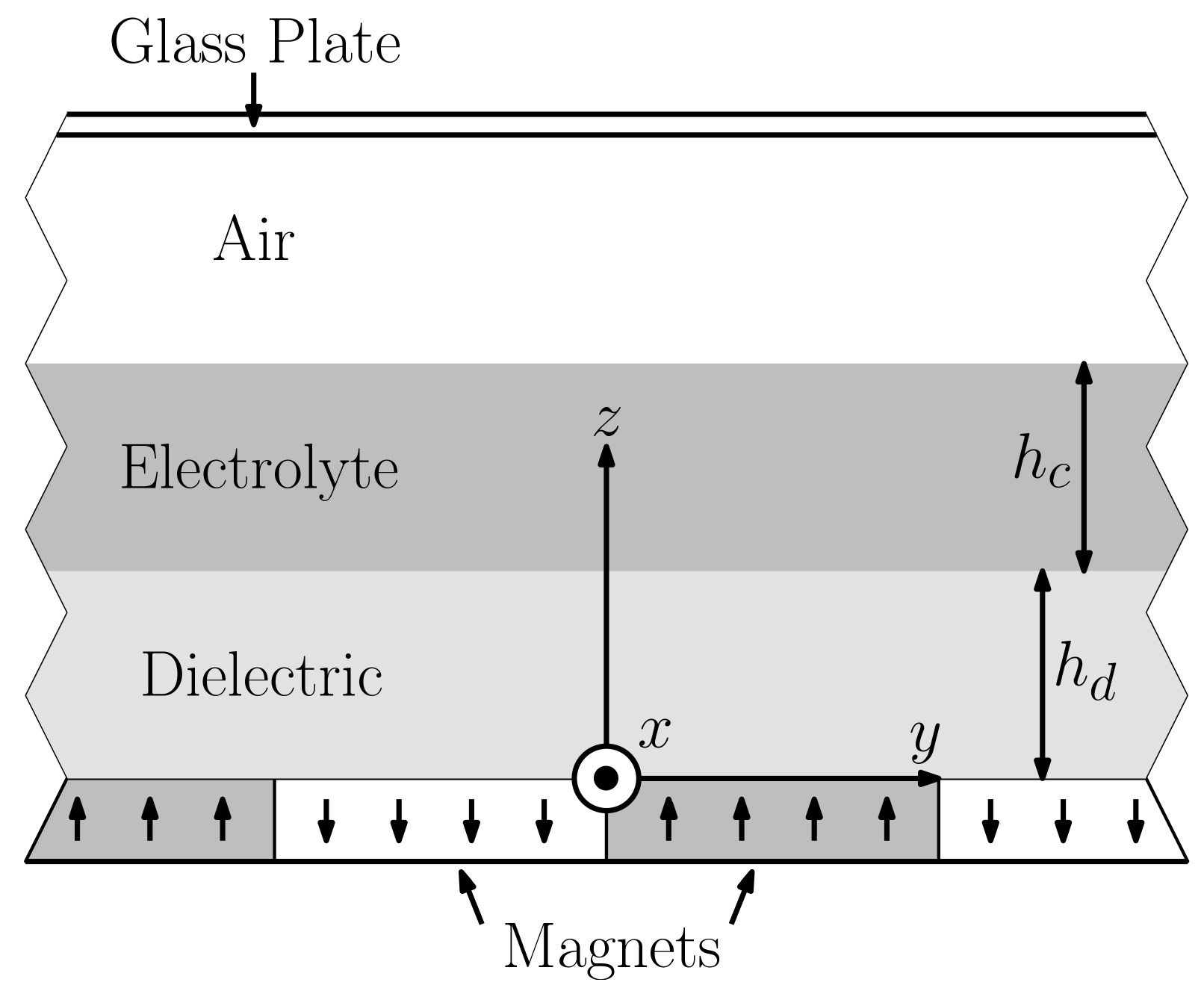}}}
\caption{\label{expsetup} The experimental setup for quasi-2D Kolmogorov-like flow, viewed (a) from above and (b) from the side.  As a uniform, steady current with density J flows between the two electrodes through the electrolyte layer, shear flows in the $\pm x$-direction (represented by arrows) are caused by spatially alternating Lorentz forces.}
\end{figure}

As mentioned earlier, Q2D flows in shallow layers of electrolytes have been realized experimentally in homogeneous as well as stratified electrolytic layers. Numerical studies \citep{satijn_2001} comparing the vertical velocity components in both these configurations have suggested that stratification does contribute to suppression of 3D motion. Hence, we use an  experimental setup which closely resembles the one suggested by \citet{rivera_2005} -- the two-immiscible-layer configuration with the electrolyte being the top layer. The setup, shown in Figure \ref{expsetup}, consists of an array of magnets placed at the center of an acrylic box of dimensions 25.4 cm$\times$20.3 cm$\times$3.8 cm. The top surface of the magnet array corresponds to the plane $z=0$. The region $0<z<h_d$ is filled with perfluorooctane, a dielectric fluid of viscosity $\mu_d = 1.30$ mPa$\cdot$s and density $\rho_d = 1769$ kg/m$^3$; above it is a layer of conducting fluid (electrolyte) of thickness $h_c$ ($h_d<z<h_d + h_c$).  We typically use volumes of fluids such that $h_d=h_c=0.3\pm0.01$ cm.  For the electrolyte, we use either of the following: a ``low viscosity electrolyte" consisting of a 0.3 M solution of CuSO$_4$ (with viscosity $\mu_c = 1.12$ mPa$\cdot$s and density $\rho_c = 1045$ kg/m$^3$) or a ``high viscosity electrolyte" consisting of a 0.3 M solution of CuSO$_4$ with 50\% glycerol by weight (with viscosity $\mu_c = 6.06$ mPa$\cdot$s and density $\rho_c = 1160$ kg/m$^3$).  The electrolyte and dielectric fluids are completely immiscible and their relative densities maintain this configuration.  A small amount of surfactant (dish soap) is added to the electrolyte to decrease the surface tension, and a glass plate is placed on top of the box to limit evaporation. A steady current is driven through the electrolyte via two 24.1 cm$\times$0.3 cm$\times$0.6 cm copper electrodes fixed along the longitudinal boundaries on either side of the box. The Lorentz forces acting on the electrolyte set the fluid layers in motion. 
\newsavebox{\tempbox}
%\begin{figure}[t!]
%\sbox{\tempbox}{\includegraphics[width=2.5in]{Figure_1_a.pdf}}
%\subfloat[]{\usebox{\tempbox}\label{}}%
%\qquad
%\subfloat[]{\vbox to \ht\tempbox{
%  \vfil
%\includegraphics[width=2.5in]{Figure_1_b.pdf}
%  \vfil}\label{}}%
%  \caption{The experimental setup for quasi-2D Kolmogorov-like flow, viewed (a) from above and (b) from the side.  As a uniform, steady current with density J flows between the two electrodes through the electrolyte layer, shear flows in the $\pm x$-direction (represented by arrows) are caused by spatially alternating Lorentz forces.}\label{expsetup}
%\end{figure}

\begin{figure}[t!]
\centerline{\subfloat[]{\includegraphics[width=2.5in]{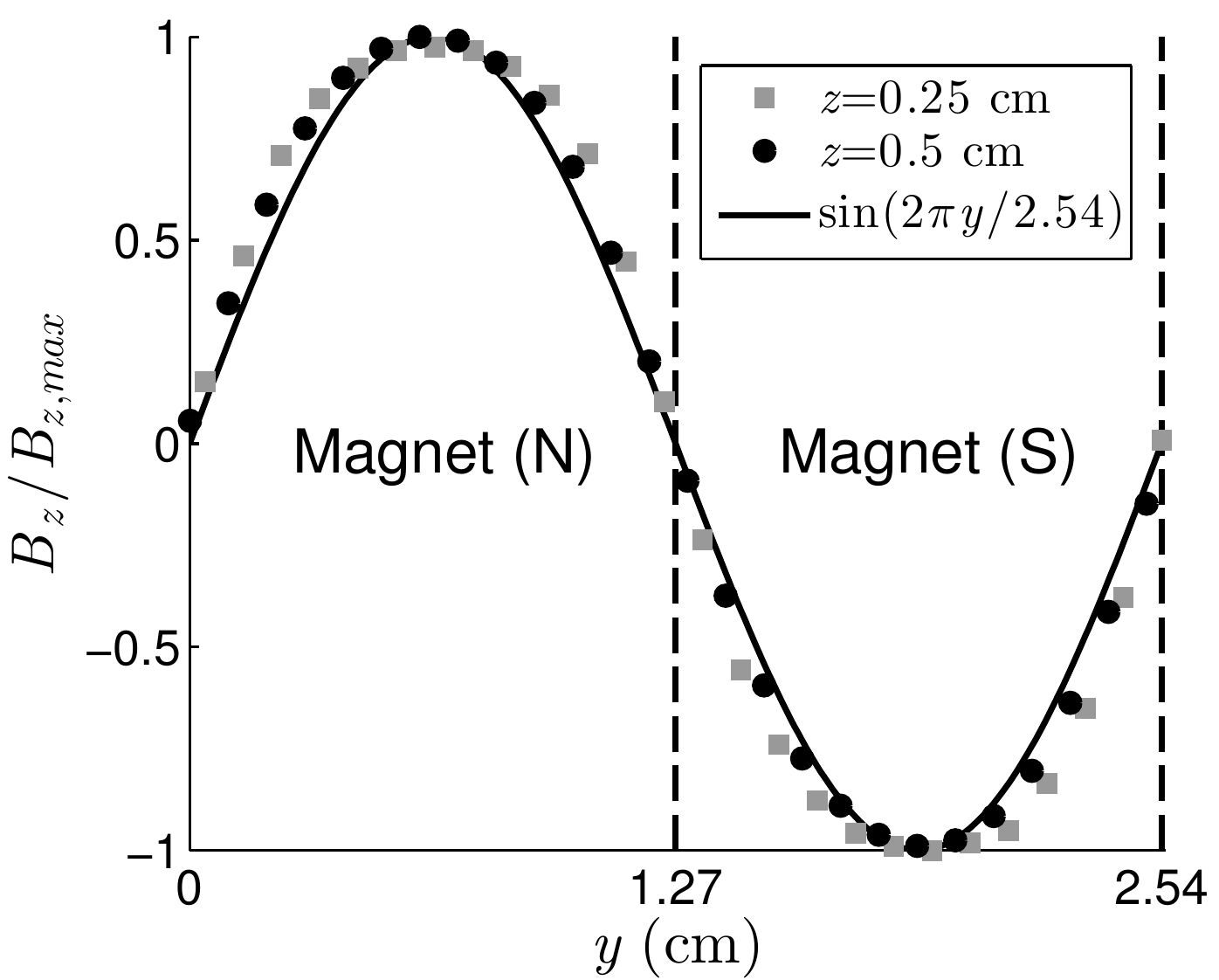}}
\subfloat[]{\includegraphics[width=2.54in]{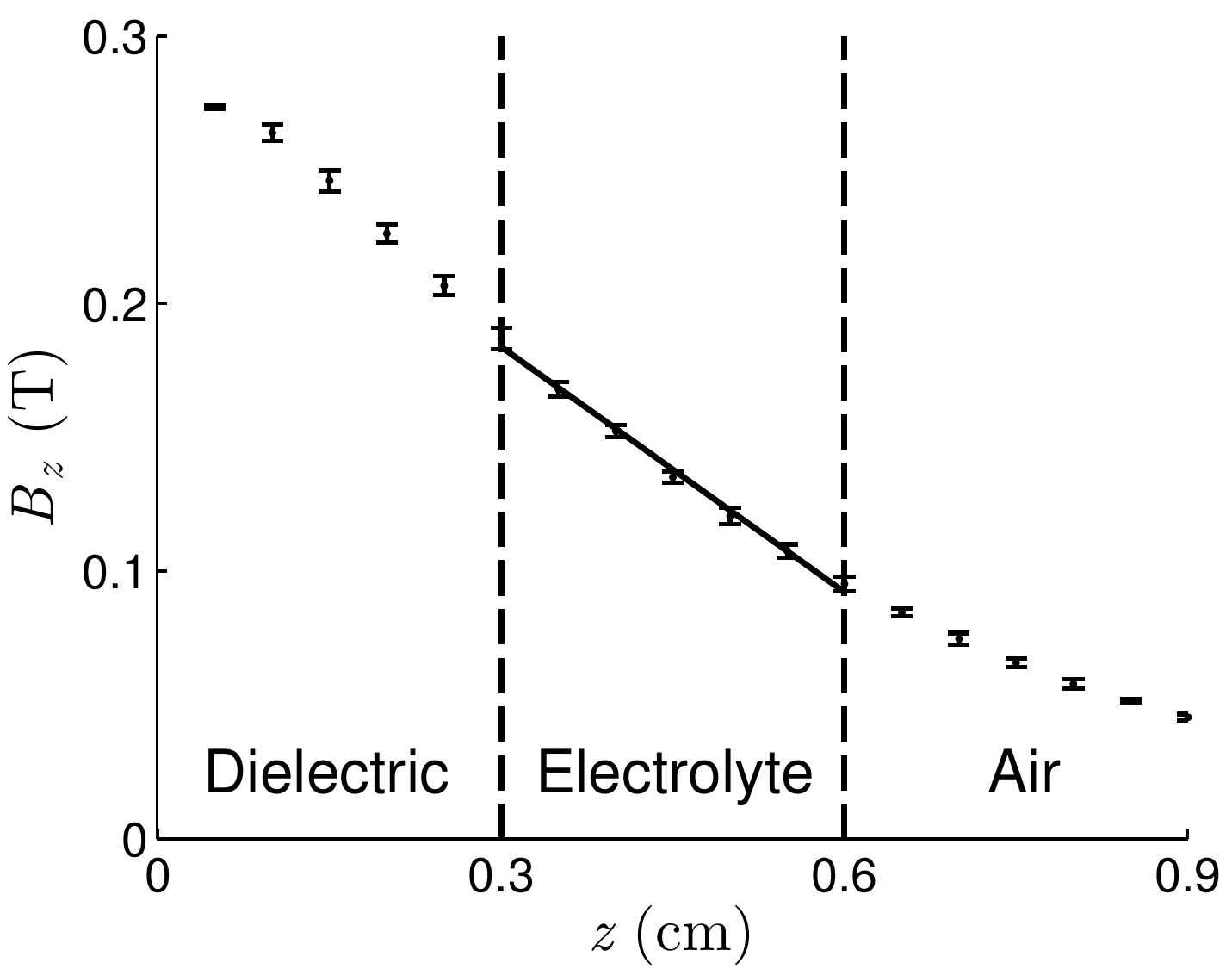}}}
\caption{\label{fig:magfield} (a) Experimental measurements of the transverse variation of the $z$-component of the magnetic field, $B_z$, above the middle two magnets of the magnet array, normalized by the maximum value of $B_z$ for fixed $z$, $B_{z,max}$.  A sine wave with periodicity equal to the width of one magnet pair is shown for comparison. (b) Experimental measurements of the decay of $B_z$ with increasing height ($z$) from the magnets' surface.  Error bars indicate one standard deviation.}
\label{fig:magfield_exp}
\end{figure}

To create a spatially periodic magnetic field, we construct a magnet array with 14 NdFeB magnets (Grade N42), each 15.2 cm long, 1.27 cm wide, and 0.32 cm thick. The magnetization is in the vertical ($z$) direction, with a surface field strength close to 0.3 T. The magnets are positioned side-by-side along their width to form a $15.2$ cm $\times$ ($14\times1.27$ cm) $\times$  $0.32$  cm array such that adjacent magnets have fields pointing in opposite directions, along the $z$-axis. The resulting net magnetic field ${\bf B}(y,z)$, close to the surface of the magnets, is quite complicated.  However, experimental measurements (using a F. W. Bell Model 6010 Gaussmeter) show that the profile for the $z$-component of the magnetic field, $B_z$, is approximately sinusoidal in $y$ beyond a height of $z=0.25$ cm (the measurements for the center pair of magnets are shown in Figure \ref{fig:magfield}(a)).
The spatial period of the magnetic field sets the horizontal lengthscale $L=$ 2.54 cm, which is large compared with the thickness $h_c+h_d\approx$ 0.6 cm of the fluid layers, as we assumed previously.
Furthermore, we find $B_z$ above the magnets to decay approximately linearly with $z$ within the typical position of the electrolyte layer (0.3 cm $\leq\! z\! \leq $ 0.6 cm) (see Figure \ref{fig:magfield}(b)); the parameters we find for the fit $B_z=B_1z+B_0$ (at the maximum in $y$) are $B_1=-30.6 \pm 0.5$ T/m and $B_0=0.276 \pm 0.01$ T.

For flow visualization, tracer particles are added to the fluid and illuminated with blue light-emitting diodes; the flow is then imaged at 7.5 frames per second.  The velocity fields are extracted by performing particle image velocimetry (PIV) using the Open Source Image Velocimetry software package (available at http://osiv.sourceforge.net/).  By separately using two different types of particles with different densities, we are able to perform PIV at either the top or the bottom surface (dielectric-electrolyte interface) of the electrolyte.  We use Glass Bubbles (K15) manufactured by 3M, sieved to obtain particles with mean radius $r=24.5\pm2$ $\mu$m and mean density $\rho \approx 150$ kg/m$^3$ for seeding the top of the electrolyte. For seeding the dielectric-electrolyte interface, we use Soda Lime Solid Glass Microspheres manufactured by Cospheric with mean radius $r=38\pm4$ $\mu$m and mean density $\rho = 2520$ kg/m$^3$. The soda lime microspheres, though denser than the dielectric fluid, stay trapped between the dielectric and electrolyte layers due to interfacial tension. This facilitates measurement of the velocity at the interface.

\section{Velocity Profile in the two-immiscible-layer setup}\label{sec:profile}

To solve for the profile $P(z)$ in (\ref{eq:2dtype}) within the two immiscible layers described in the experiment, we assume that the magnet array is infinitely long in the $x$ (longitudinal) direction and periodic in the $y$ (transverse) direction. By symmetry, the magnets produce a magnetic field that has no component along the longitudinal direction:
\begin{equation}
{\bf{B}} = B_y(y,z) {\bf{\hat{y}}}+ B_z(y,z) {\bf{\hat{z}}}.
\end{equation}

Since the $z$-component of the magnetic field in the electrolyte varies linearly with $z$ and roughly sinusoidally with $y$ (cf. Figure \ref{fig:magfield_exp}), we can write
\begin{equation}\label{eq:magfield}
B_z = (B_1z+B_0)\sin(\kappa y),
\end{equation}
where $\kappa=\pi/w$ and $w$ is the width of each magnet. A uniform and constant current with density ${\bf{J}} = J{\bf{\hat{y}}}$ passing through the electrolyte along the tranverse direction results in the Lorentz force density which is given by
\begin{equation}\label{eq:forcing}
{\bf{f}} = {\bf{J}}\times{\bf{B}} = \left\{
\begin{array}{ll}
J(B_1z+B_0)\sin(\kappa y){\bf{\hat{x}}}, & h_d<z<h_d+h_c, \\
0, & 0<z<h_d \end{array}
\right.
\end{equation}
in the electrolyte and the dielectric, respectively.
 
For small current density $J$, the direction of the horizontal flow profile ${\bf{U}}$ follows that of the forcing (\ref{eq:forcing}), so we can look for laminar solutions of the form
\begin{equation}\label{eq:laminar}
{\bf{U}}(x,y,t) = u_0 \sin(\kappa y) {\bf{\hat{x}}}.
\end{equation}  
Substituting this into (\ref{eq:genprof}) yields a hydrostatic pressure distribution and a boundary value problem for the vertical profile
\begin{align}
P'' - \kappa^2P &= - \frac{J}{u_0\mu_c} (B_1z + B_0),
&h_d < z < h_d + h_c,\\
P'' - \kappa^2P &= 0, &0 < z < h_d
\end{align}
where the prime denotes differentiation with respect to $z$. The solutions to the above differential equations are given by
\begin{equation}\label{eq:2dvp}
P_\kappa = \left\{
\begin{array}{ll}
Ce^{\kappa z}+De^{-\kappa z}+\frac{JB_1}{u_0\mu_c\kappa^2} z+ \frac{JB_0}{u_0\mu_c\kappa^2}, & h_d<z<h_d+h_c, \\
Ee^{\kappa z}+Fe^{-\kappa z}, &0<z<h_d. \end{array}
\right.
\end{equation}

The unknown coefficients $C$, $D$, $E$, and $F$ can be obtained using the continuity of the velocity and stress at the dielectric-electrolyte interface ($z = h_d$), the no-slip boundary condition at the bottom of the dielectric ($z = 0$), and the stress-free boundary condition at the top of the electrolyte ($z = h_d+h_c$, electrolyte-air interface):
\begin{equation}\label{eq:bc}
\mu_d P'(h_d^-) = \mu_c P'(h_d^+), \quad P(h_d^-) = P(h_d^+), \quad P(0) = 0, \quad  P'(h_d + h_c) = 0. 
\end{equation}
Finally, $u_0$ can be found using (\ref{eq:norm}).

The factors that are mainly responsible for the inherent three-dimensionality of Q2D flows are the inhomogenity in the magnetic field and the no-slip boundary condition. For a two-immiscible-layer setup we can use the ratio of velocities at the top and bottom of the electrolyte layer as a measure that characterizes this inherent three-dimensionality:
\begin{equation}\label{eq:ratgen}
s = \frac{P(h_d+h_c)}{P(h_d)}.
\end{equation}
For a monotonically varying profile, the value of $s$ describes how strongly the magnitude of the velocity varies with $z$ in the electrolyte, with $s=1$ corresponding to a $z$-independent velocity profile.  Unfortunately, the functional form of expression (\ref{eq:ratgen}) for this flow is quite unwieldy and does not allow one to easily deduce the dependence on experimental parameters. Furthermore, closed form expressions for the coefficients in the modified 2D vorticity equation (\ref{eq:2dterms}) also turn out to be too complicated to yield much insight.

\begin{subsection}{Unidirectional flow}
  
We can derive a relatively simple analytical expression for the ratio of velocities $s$ in the special case where we ignore the $y$ dependence of the magnetic field $B_z$, which can be thought of as the $\kappa\to0$ limit of (\ref{eq:magfield}). In this case $B_z = B_1z + B_0$, and the laminar flow is given by ${\bf{U}}(x,y,t) = u_0 {\bf{\hat{x}}}$. The solution (\ref{eq:2dvp}) is then replaced by 
\begin{equation}\label{eq:1dvp}
P_0 = \left\{
\begin{array}{ll}
- \frac{JB_1}{6u_0\mu_c} z^3 - \frac{JB_0}{2u_0\mu_c} z^2 + Cz + D, & h_d<z<h_d+h_c, \\
Ez+F, & 0<z<h_d. \end{array}
\right.
\end{equation}

Although the functional forms (\ref{eq:2dvp}) and (\ref{eq:1dvp}) of the velocity profile are quite different for the Kolmogorov flow and the uniform flow, their shape is virtually indistinguishable, as Figure \ref{fig:vel_prof} illustrates. This suggests that Q2D flows with arbitrary horizontal flow profiles ${\bf{U}}(x,y,t)$ and moderately high Reynolds numbers may be accurately described using the simple velocity profile (\ref{eq:1dvp}).

Computing the coefficients using the boundary conditions (\ref{eq:bc}) we find that the ratio of the velocities at the top and the bottom of the electrolyte layer is given by
\begin{equation}\label{eq:vratio}
s = 1 + \frac{1}{2}\frac{\mu_dh_c}{\mu_ch_d}\left(1 + \frac{1}{6}\frac{\triangle B}{B_{mean}}\right),
\end{equation}
where $\triangle B = B_1h_c$ is the change in magnetic field across the electrolyte and $B_{mean} = B_0 + B_1h_b + \frac{1}{2}B_1h_c$ is the mean magnetic field in the electrolyte.
 
The coefficients (\ref{eq:2dterms}) that appear in the modified 2D vorticity equation (\ref{eq:2dvor}), in addition to depending explicitly on the experimental parameters, depend on the profile $P(z)$ as well. Since the shapes of the profiles $P_0(z)$ and $P_{\kappa}(z)$ are virtually indistinguishable, we can evaluate analytical expressions for the coefficients (\ref{eq:2dterms}) using $P_0(z)$. For the Rayleigh friction, using the velocity profile $P_0(z)$ we obtain:
\begin{equation}\label{eq:rayfric}
\alpha = \frac{\frac{\mu_d}{\rho_c}\frac{1}{h_d h_c}}{1 + \frac{1}{2}\frac{h_d}{h_c}\frac{\rho_d}{\rho_c} + \frac{1}{3}\frac{\mu_d}{\mu_c}\frac{h_c}{h_d}\left(1+\frac{1}{8}\frac{\triangle B}{B_{mean}}\right)}.
\end{equation}
For the set of parameters used in the experiment, the Rayleigh friction coefficient has a very weak dependence on $B_1$ and $B_0$, with the term containing $\slfrac{\triangle B}{B_{mean}}$ contributing approximately $2\%$ to the friction coefficient, so for practical purposes one can set $\triangle B=0$.

For the depth-averaged kinematic viscosity, we obtain: 
\begin{equation}\label{eq:nu}
%\nu = \nu_c\frac{\frac{1}{3}+\frac{1}{24}\frac{\triangle B}{B_{mean}}+ \frac{h_d}{h_c}\frac{\mu_c}{\mu_d}+\frac{1}{2}\left(\frac{h_d}{h_c}\right)^2}{\frac{1}{3}+\frac{1}{24}\frac{\triangle B}{B_{mean}}+ \frac{h_d}{h_c}\frac{\mu_c}{\mu_d}+\frac{1}{2}\frac{\mu_c}{\mu_d}\frac{\rho_d}{\rho_c}\left(\frac{h_d}{h_c}\right)^2},
\nu = \nu_c\frac{1 + \frac{1}{2}\frac{h_d}{h_c}\frac{\mu_d}{\mu_c} + \frac{1}{3}\frac{\mu_d}{\mu_c}\frac{h_c}{h_d}\left(1+\frac{1}{8}\frac{\triangle B}{B_{mean}}\right)}{1 + \frac{1}{2}\frac{h_d}{h_c}\frac{\rho_d}{\rho_c} + \frac{1}{3}\frac{\mu_d}{\mu_c}\frac{h_c}{h_d}\left(1+\frac{1}{8}\frac{\triangle B}{B_{mean}}\right)},
\end{equation}
where $\nu_c$ is the kinematic viscosity of the conducting layer.

\begin{figure}[t!]
\centering
\subfloat[]{\includegraphics[width=2.5in]{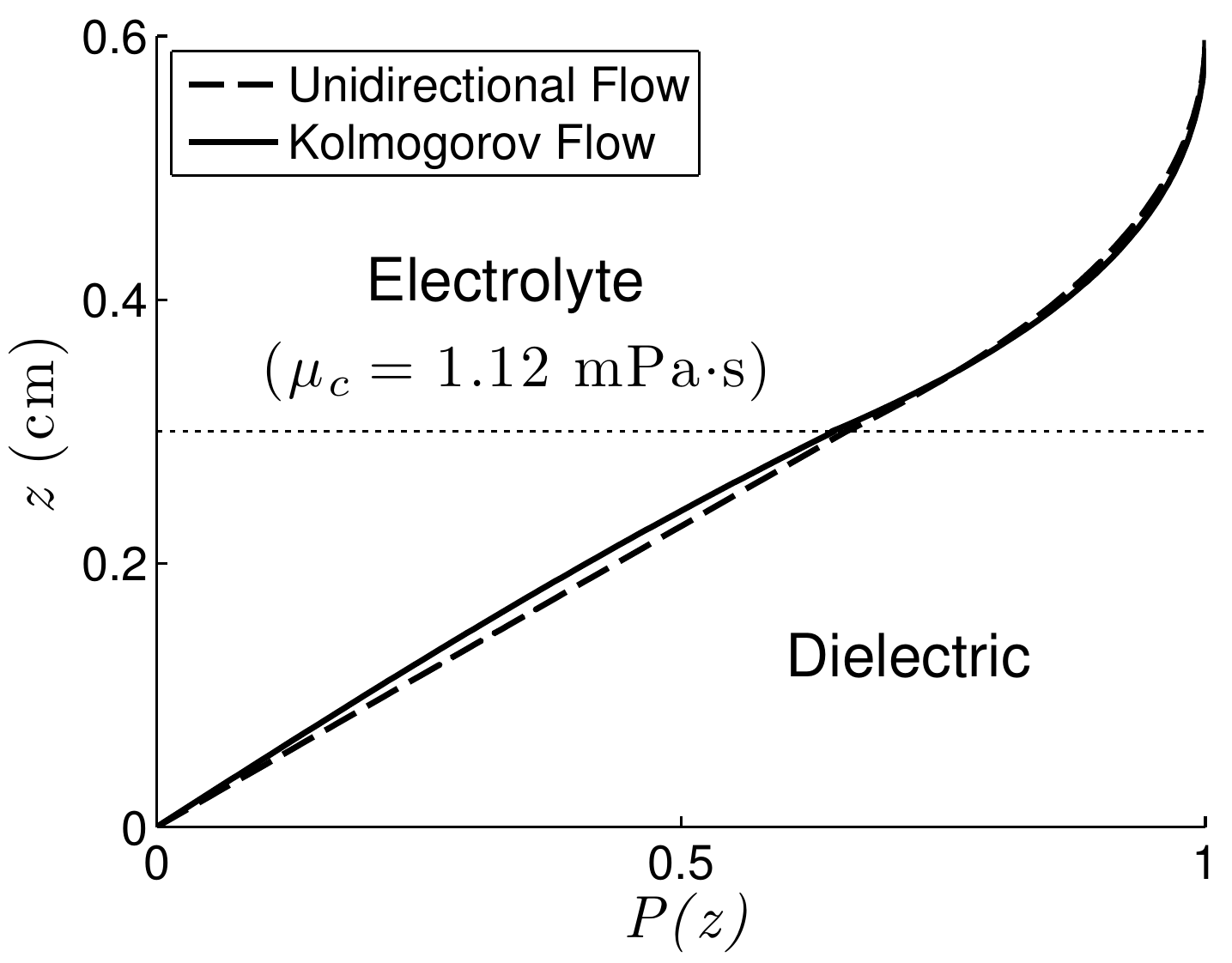}}
\subfloat[]{\includegraphics[width=2.31in]{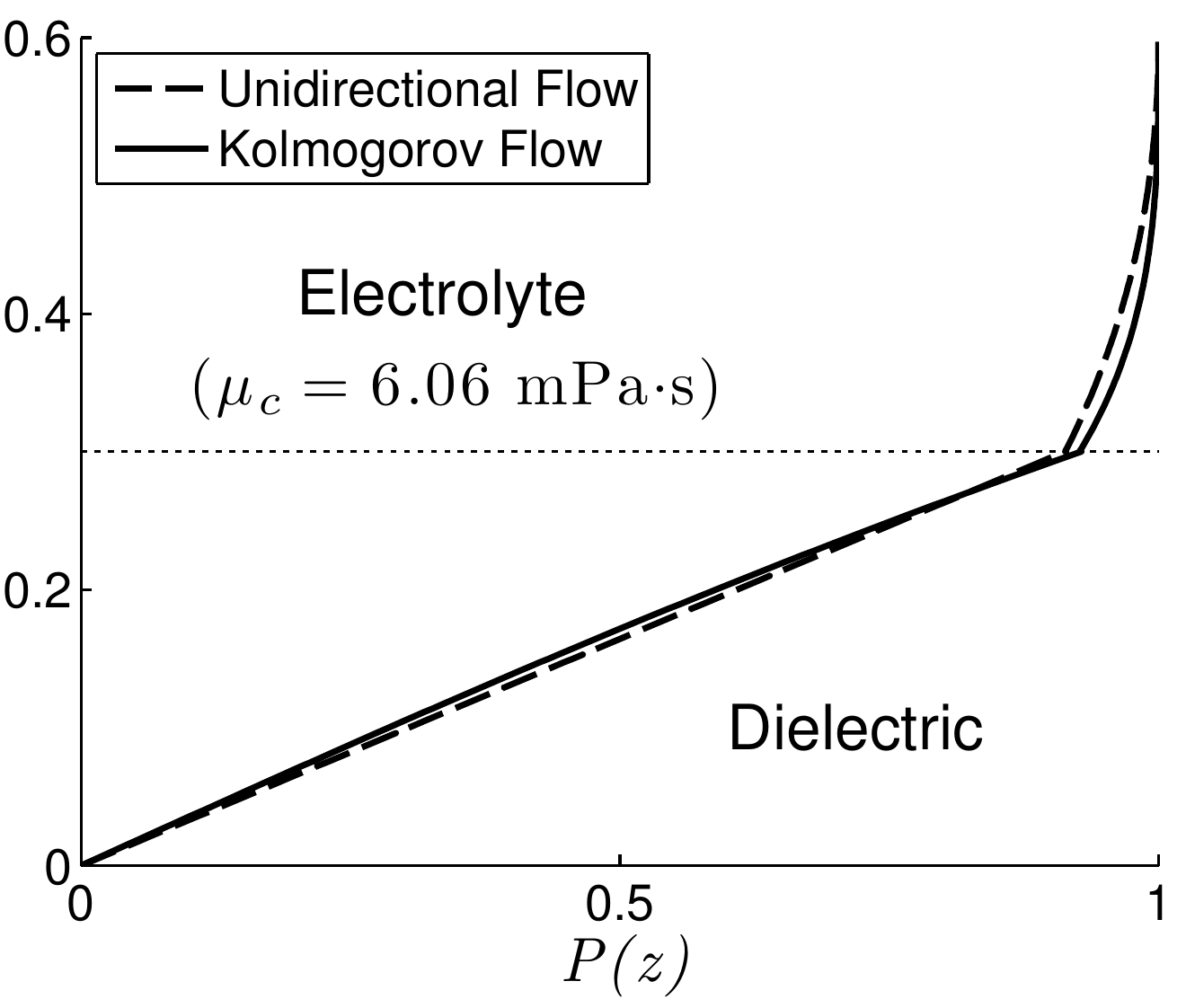}}
\caption{\label{fig:vel_prof} Analytical results for the vertical flow profile in both layers, with $h_d=h_c=0.3$ cm, for (a) the low viscosity electrolyte and (b) the high viscosity electrolyte.  The ratios of the velocities, as defined by (\ref{eq:ratgen}), are: (a) uniform flow: $s_{low} = 1.52$, Kolmogorov flow: $s_{low} = 1.55$ and (b) uniform flow: $s_{high} = 1.09$, Kolmogorov flow: $s_{high} = 1.08$.}
\end{figure}

\end{subsection}

\begin{figure}
\centering
\subfloat[]{\includegraphics[width=2.5in]{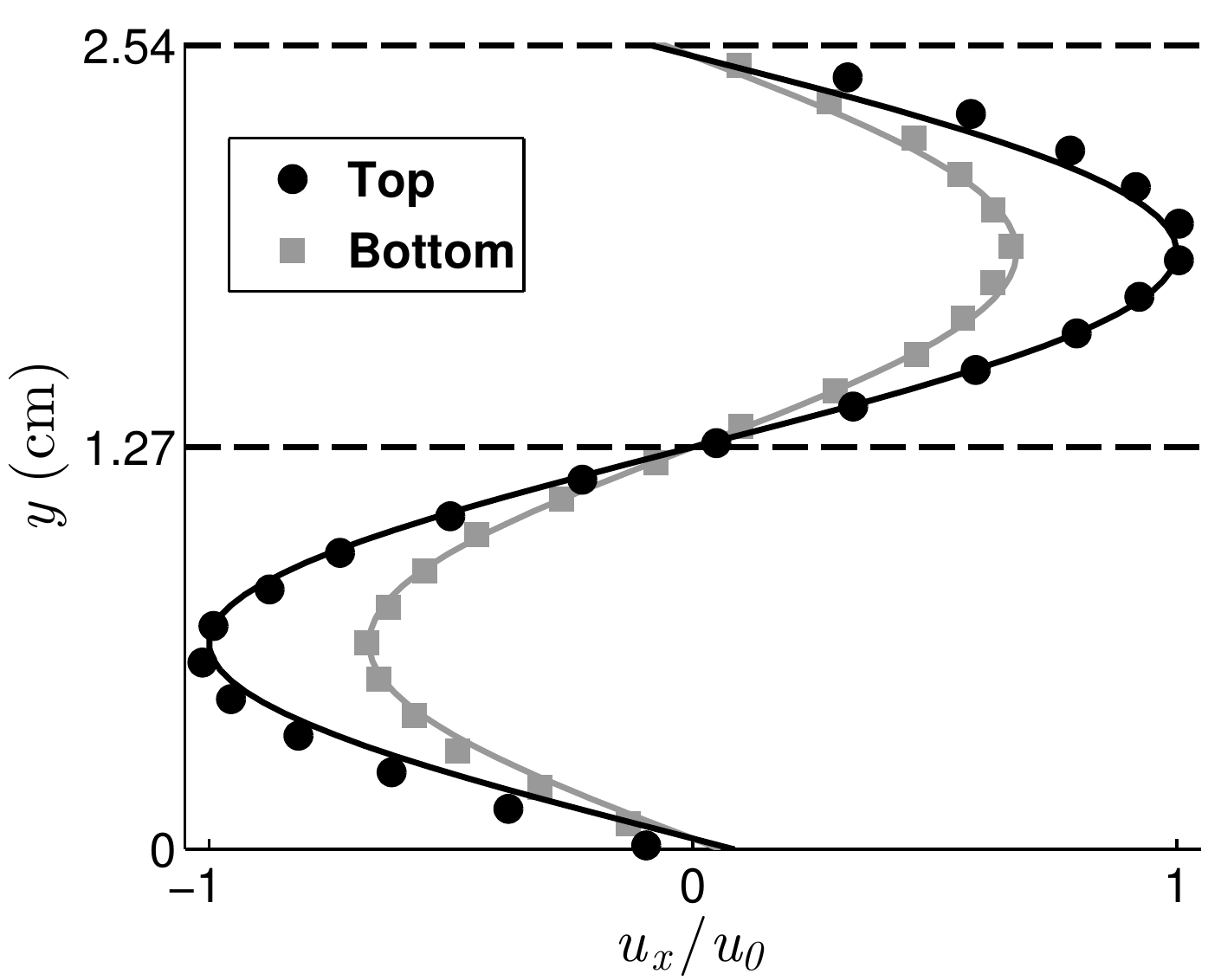}}
\subfloat[]{\includegraphics[width=2.4in]{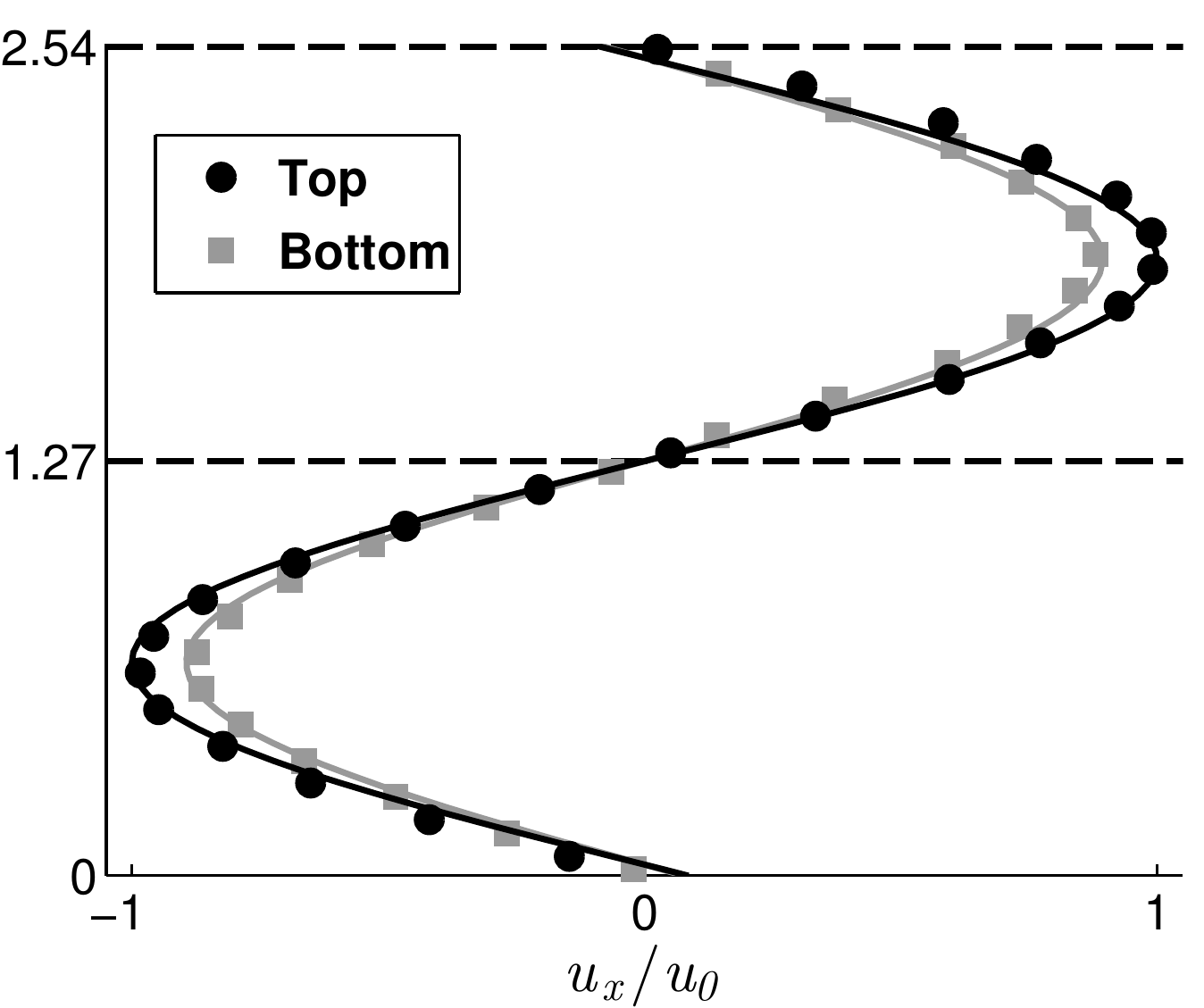}}
\caption{Experimental measurements of the horizontal flow profile in the electrolyte layer, with $h_d=h_c=0.3$ cm, for (a) the low viscosity electrolyte and (b) the high viscosity electrolyte.  PIV measurements are plotted for the time-independent laminar flow near the center of the magnet array; each data point is time-averaged over 5 minutes and spatially averaged over 4.5 cm along the $x$-direction to obtain an accurate estimate of the mean.  A sine wave with fixed periodicity is fit to each data set, and the velocities are normalized by the amplitude of the top layer fit, $u_0$. Errorbars are the size of the symbols.}
\label{fig:vel_cs}
\end{figure}

\section{Results}

\subsection{Enhanced two-dimensionality in the electrolyte}

Expression (\ref{eq:vratio}) suggests that the shallower the electrolyte layer is (relative to the dielectric layer), the closer one comes to a vertically uniform profile in the electrolyte ($s = 1$). However, electrolyte layers with thickness less than 0.25 cm are found to be unstable in the experiment, as they break open to form configurations that correspond to lower total surface energy. Alternatively, one may increase the thickness $h_d$ of the dielectric layer. However, this has the drawback that one moves farther away from the ``shallow layer'' approximation. Hence, the most straightforward way to make the flow in the electrolyte nearly two-dimensional is by increasing the ratio of viscosities. The choice of how high the viscosity of the electrolyte should be is tricky, since it becomes increasingly difficult to drive flows in a fluid with very high viscosity. Hence we choose a 10\% limit ($s = 1.1$), rather arbitrarily, on how much the velocity at the top and bottom of the electrolyte layer should differ by. Using $\mu_d = 1.30$ mPa$\cdot$s, $\slfrac{\triangle B}{B_{mean} \approx 0.6}$ and $h_d = h_c$ we see that $\mu_c \geq 6.0$ mPa$\cdot$s for reaching this limit of $s=1.1$. Comparison of the analytical velocity profiles presented in Figure \ref{fig:vel_prof} shows that the uniformity of the velocity in the conducting layer is substantially enhanced when a more viscous electrolyte is used. This is confirmed (see Figure \ref{fig:vel_cs}) by the measurements of the in-plane velocity profile of the laminar flow at the top, as well as at the bottom, of the electrolyte. We find that the flow in the high viscosity electrolyte is much closer to being vertically uniform ($s_\mathrm{high}\approx 1.08$) than in the low viscosity electrolyte ($s_\mathrm{low} \approx 1.50$).  

\subsection{Comparison between theory and experiment}

\begin{figure}[t]
\centering
\subfloat[]{\includegraphics[height=2.2in]{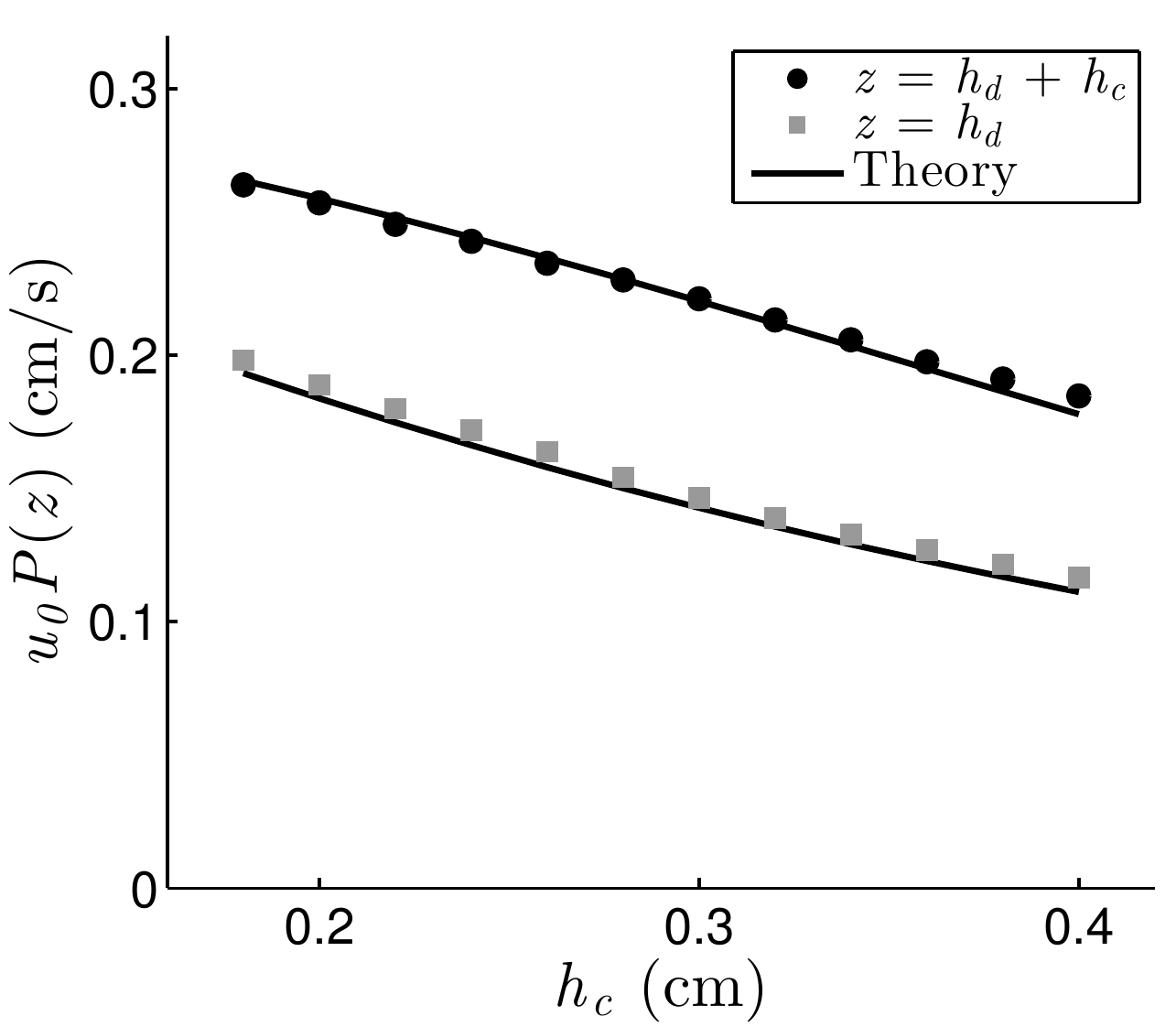}}\hspace{5mm} 
\subfloat[]{\includegraphics[height=2.2in]{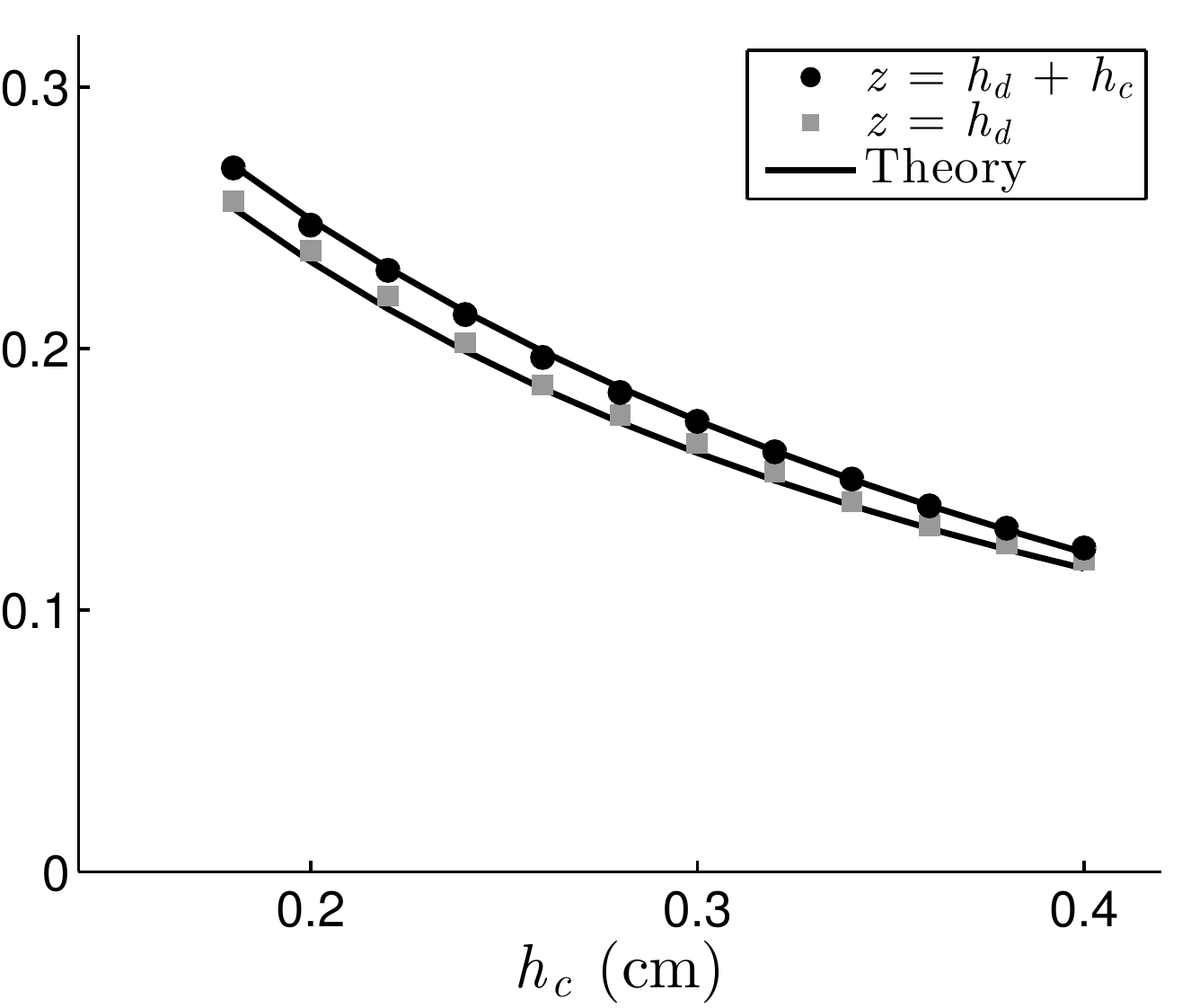}}
\caption{\label{fig:expandtheory} Comparison of experimental and theoretical results for $u_0 P_\kappa(h_c+h_d)$ and $u_0P_\kappa(h_d)$, which correspond to the amplitude of the sinusoidal velocity profile at the top and the bottom of the electrolyte layer, respectively. Here, $h_c$ is varied while $h_d=0.3$ cm is held constant.  Plots correspond to (a) the low viscosity electrolyte (with constant current $I$ = 2.1 mA) and (b) the high viscosity electrolyte (with constant current $I$ = 5.0 mA).  PIV measurements of the time-independent laminar flow are time-averaged over 5 minutes to reduce experimental noise and then fit with a sine wave with fixed periodicity. Errorbars are the size of the symbols.}
\end{figure}

Figure \ref{fig:expandtheory} shows experimental measurements of the velocity amplitude of the laminar flow at the top and the bottom of the electrolyte as the thickness $h_c$ of the electrolyte layer is varied, while keeping the current $I$ constant. Also plotted for comparison are the theoretical predictions of $u_0P_\kappa(h_d+h_c)=u_0$ and $u_0P_\kappa(h_d)=u_0/s$, which denote the velocity at the top and the bottom of the electrolyte, respectively. Most importantly, all the parameters used in the theoretical calculations have been measured  experimentally. As can be seen from the plots, the relative difference between theory and experiment, for electrolytes of both viscosities, is less than 5\%. This provides an indirect validation of the shape $P_{\kappa}$ of the velocity profile shown in Figure \ref{fig:vel_prof}.

\subsection{Coefficients in the generalized 2D vorticity equation}

The motivation behind estimating the shape of the profile $P(z)$ in the two-immiscible-layer setup was, in part, to determine the coefficients (\ref{eq:2dterms}) that appear in the generalized 2D vorticity equation (\ref{eq:2dvor}). For the low viscosity electrolyte ($\mu_c = 1.12$ mPa$\cdot$s), we obtain $\beta = 0.72$, $\nu = 0.94$ cSt, and $\alpha = 0.063$ s$^{-1}$. This estimate of the friction coefficient is a factor of two smaller than the one suggested by \citet{rivera_2005}. For the high viscosity electrolyte ($\mu_c = 6.06$ mPa$\cdot$s), we obtain $\beta = 0.82$, $\nu = 3.35$ cSt, and $\alpha = 0.068$ s$^{-1}$. It is important to note that the friction coefficient remains fairly insensitive to the viscosity of the electrolyte. This has a significant consequence that one can change the relative importance of the diffusion term $\nu \nabla^2_\parallel \Omega$ and the friction term $-\alpha \Omega$ in the vorticity equation (\ref{eq:2dvor}) by changing the viscosity of the upper layer in the experiment. Interestingly, the friction coefficient for the two-immiscible-layer system is {\it not} very different from the one computed for a homogeneous layer of fluid. Using a Poiseuille-like profile for $P(z)$ and choosing $h = h_c + h_d$ and $\nu \approx 1$ cSt we obtain $\alpha = \slfrac{\pi^2 \nu}{4 h^2} \approx 0.07$ s$^{-1}$.  
 
\subsection{Spin-down comparison}

After the forcing is switched off, $W = 0$, the flow decays to rest exponentially fast, dissipating energy via bottom friction ($-\alpha \Omega$) as well as diffusion ($\nu \nabla^2_\parallel \Omega$). The solution of (\ref{eq:2dvor}) corresponding to the initial condition (\ref{eq:laminar}) describing Kolmogorov flow is $\Omega(x,y,t) = \Omega_0\exp(-t/\tau) \cos(\kappa y)$, where the decay rate is given by
\begin{equation}\label{eq:spindown}
\tau^{-1} = \alpha + \kappa^2 \nu.
\end{equation}

As a check of our parameter estimation in the 2D model, we compare the temporal evolution of the flow by letting it decay to rest from the steady laminar flow (\ref{eq:laminar}) after suddenly turning the current off. Experimentally, we observe that, after a brief transient, the velocity profile ${\bf{U}}(x,y,t) = u_0(t)\sin(\kappa y)$ measured at the top of the electrolyte decays exponentially, $u_0(t)\sim \exp(-t/\tau)$. These measurements yield a decay rate of $\tau^{-1}_\mathrm{low}=0.14 \pm 0.01$ s$^{-1}$ for the low viscosity electrolyte and $\tau^{-1}_\mathrm{high}=0.3 \pm 0.01$ s$^{-1}$ for the high viscosity one. In comparison, the analytical solution (\ref{eq:spindown}) yields  $\tau^{-1}_\mathrm{low}=0.12 \pm 0.007$ s$^{-1}$ for the low viscosity electrolyte and $\tau^{-1}_\mathrm{high}=0.29 \pm 0.009$ s$^{-1}$ for the high viscosity one. It is important to note that equation (\ref{eq:spindown}) does not account for the change in the shape of the profile $P(z)$ during the decay, which likely explains the slight disagreement between the theory and experiment at low  $\mu_c$. However, as one can see, in the presence of temporal variation, we achieve significantly better agreement with the theory for the high viscosity electrolyte than for the low viscosity one. This is a non-trivial result: although the flow in the high viscosity electrolyte is very nearly two-dimensional, the flow in the dielectric never is.

\section{Conclusion}

Recent studies aimed at understanding 2D turbulence from a dynamical systems perspective have found an abundance of exact but unstable solutions of the NSE at low Reynolds numbers. At these Reynolds numbers, flows realized experimentally in shallow electrolytic layers are known to be Q2D. In this article, assuming Q2D behavior of the flow, we depth-average the NSE and derive a generalized 2D vorticity equation. The friction term appears naturally as a consequence of the no-slip boundary condition at bottom. Furthermore, we have shown the presence of a prefactor to the advection term, necessary for a quantitative comparison between 2D DNS and experimental measurements  at the top surface. Using the Kolmogorov flow and the unidirectional flow models, we have derived the velocity profiles along the vertical direction in the two-immiscible-layer system and have shown that they are essentially indistinguishable, suggesting that, at moderate $Re$, a universal profile can be used for these and other flows.
  
We quantify the inherent three-dimensionality of the flow using the ratio of velocity at the free surface to that at the interface. Using this measure, we see that increasing the viscosity of the fluid in the upper layer with respect to the one in the bottom layer makes the flow in the former closer to uniform, which is advantageous for a number of reasons: (i) We find better agreement with the 2D model for both steady and decaying laminar flow--which {\it{may}} extend into time-dependent regimes, (ii) The flow in the electrolyte is closer to being two-dimensional, in the sense that the velocity dependence on $z$ is greatly reduced, and (iii) The vertical velocity component is reduced due to smaller gradients in vorticity across the layer, which should suppress Eckman pumping within the electrolyte. One potential drawback of increasing the ratio of viscosities is the possibility of the less viscous bottom layer going turbulent before the upper layer does \citep{akkermans_2010}. To address the possibility of excess Joule heating, experiments conducted using the electrolyte with higher viscosity, while forcing the flow steadily for 60 minutes at $Re \approx 40$, have shown that the fluid temperature increases only by around 1 $^\circ$C. Hence in the regime of interest, one could ignore the effects of Joule heating. 

We confine the comparison between theory and experiment to the laminar flow because we seek closed form expressions for the coefficients in the 2D vorticity equation, to gain insight into how they depend on various experimental parameters. A logical extension of this study would be a comparison in the forced time-dependent regime, which would require numerical simulations with boundary conditions mimicking those in the experiment -- at least for spatially extended flows like the Kolmogorov flow. As an alternative, one could attempt such a comparison using a dipolar vortex \citep{figuearoa_2009} or a periodic lattice of vortices \citep{juttner_1997}, with the magnetic field being significantly more complicated in the former case.

\section{Acknowledgments}
We thank Daniel Borrero for his valuable suggestions and insights on both experimental and theoretical aspects of the problem. We also thank Phillip First for lending us the Gaussmeter from his laboratory. This work is supported in part by the National Science Foundation under grants No. CBET-0853691, CBET-0900018, and CMMI-1234436. 

\bibliographystyle{apsrev-title}
\bibliography{velocity_profile_in_a_two_layer_kolmogorov_like_flow}
\end{document}